\documentclass[aps,preprint]{revtex4}
\usepackage{graphicx}

\begin{document}

\title{A semiclassical model of light mesons}
\author{F. Brau
and C. Semay
}
\address{Universit\'{e} de Mons-Hainaut, Place du Parc 20,
B-7000 Mons, BELGIQUE}
\author{B. Silvestre-Brac
}
\address{Institut des Sciences Nucl\'{e}aires,
Avenue des Martyrs 53, F-38026 Grenoble-Cedex, FRANCE}
\date{\today}

\begin{abstract}
The dominantly orbital state description is applied to the study of
light mesons. The effective Hamiltonian is characterized by a
relativistic kinematics supplemented by the usual funnel potential
with a mixed scalar and vector confinement. The influence of two
different finite quark masses and potential parameters on Regge and
vibrational trajectories is discussed.
\end{abstract}
\pacs{12.39.Ki,12.39.Pn,12.40.Nn} 

\maketitle

\section{Introduction}
\label{sec:intro}

Potential models, in particular semirelativistic ones, have been
proved
extremely successful for the description of mesons and baryons
\cite{luch91}. Several technics have been developed in order to solve
numerically the spinless Salpeter equation used in most works devoted
to the study of light mesons. Variational calculations performed in
well chosen bases \cite{fulc94,brau98a}, and the three-dimensional
Fourier grid Hamiltonian method \cite{brau98b}, for instance, can
provide accurate numerical results. Nevertheless, it is interesting to
obtain analytical results as well. The dominantly orbital state (DOS)
description has been developed in this direction \cite{goeb90,olss97}.
In this approach, the orbitally excited states are obtained as a
classical result while the radially excited states can be treated
semiclassically. It is then possible to obtain information about
family of states characterized by a high orbital angular momentum.

\par In Ref.~\cite{olss97}, the DOS model has been applied to study
light-light and heavy-light meson spectra with a confinement potential
being a mixture of scalar and vector components. In the case of one or
two quarks with vanishing masses, it was shown that linear Regge
(orbital) and vibrational (radial) trajectories are obtained for an
arbitrary scalar-vector mixture but that the ratio of radial to
orbital energies is strongly dependent on the mixture. These results
have been extended in Ref.~\cite{silv98} by considering the presence
of
a quark and an antiquark with finite identical masses and the
introduction of a Coulomb-like interaction plus a constant potential
in
addition to the confinement. In the limit of small masses and small
strength of the short range potential, or in the limit of large
angular momentum, it is shown that all calculations can be worked out
analytically. One result is that slopes of orbital and radial
trajectories depend only on the string tension and on the
vector-scalar mixture in the confinement potential. Moreover, small
finite quark masses do not alter significantly the linearity,
specially in the case of dominant vector-type confining potential. As
expected, the Coulomb-like potential has no effect on trajectories but
its influence on the meson masses is determined in the approximation
of the DOS description. Lastly, it is shown that the ratio of radial
to orbital trajectory slopes depends on the scalar-vector confinement
mixture only.

\par In this work, we consider the case of a quark and an antiquark
with finite but different masses. The potential taken into account is
the same as the one considered in our previous work, that is to say a
linear scalar-vector mixed confinement potential supplemented by a
Coulomb-like interaction. An expression for the square mass of a
meson is obtained in the limit of small quark masses and small
strength of the short range potential, or in the limit of large
angular momentum. The coefficients of the square mass formula
cannot be obtained analytically but they are computed numerically.
Nevertheless a coverage of all possible situations is performed. The
square mass formula is established and discussed in
Sec.~\ref{sec:model} and Sec.~\ref{sec:disc} respectively. Some
concluding remarks are given in Sec.~\ref{sec:rem}.

\section{The model}
\label{sec:model}

\subsection{The dominantly orbital state description}
\label{ssec:dos}

A detailed explanation of the technique of the DOS description is
given in Ref.~\cite{olss97}. So we just recall the basic ideas and
we only focus on differences between our work and previous ones. In
all our formulas, we use the natural units $\hbar = c = 1$.

\par We consider a system composed of two particles with masses $m_1$
and $m_2$ interacting via a scalar potential $S(r)$ and a vector
potential $V(r)$ which depend only on the distance $r$ between the
particles. In the center-of-mass laboratory, the classical mass $M$
of the system characterized by a total orbital angular momentum $J$ is
given by
\begin{equation}
\label{mclass}
M(r,p_r,J) = \sqrt{p_r^2+\frac{J^2}{r^2}+\left( m_1+\alpha_1 S(r)
\right)^2} + \sqrt{p_r^2+\frac{J^2}{r^2}+\left( m_2+\alpha_2 S(r)
\right)^2} + V(r),
\end{equation}
where $p_r$ is the radial internal momentum. The parameters
$\alpha_1$ and $\alpha_2$ indicate how the scalar potential $S(r)$ is
shared among the two masses. These quantities must satisfy the
following conditions to ensure a good nonrelativistic limit:
\begin{equation}
\label{alpha}
\alpha_1 + \alpha_2 = 1 \nonumber \quad \text{and} \quad
\lim_{m_i \rightarrow \infty} \alpha_i = 0.
\end{equation}
A natural choice is to take
\begin{equation}
\label{alpha2}
\alpha_1 = \frac{m_2}{m_1+m_2} \quad \text{and}
\quad \alpha_2 = \frac{m_1}{m_1+m_2}.
\end{equation}
We will use this prescription in this work. In the following, it is
assumed that $m_1 \leq m_2$. A parameter $\beta = m_1/m_2$ is
introduced to measure the mass asymmetry, and the heaviest quark mass
is denoted by $m$. Contrary to the philosophy adopted in our previous
paper \cite{silv98}, it is necessary to work with the hamiltonian
formalism and to give up the lagrangian formalism.

\par The idea of the DOS description is to make a classical
approximation by
considering uniquely the classical circular orbits, that is to say the
lowest energy
states with a given $J$. This state is defined by $r=r_0(J)$, and
thus $dr/dt=0$
and $p_r=0$. Let us denote $M_0(J)=M(r=r_0,p_r=0,J)$.
In order to get the
radial excitations, a harmonic approximation around a
classical circular orbits is calculated. If the harmonic quantum
energy is given by
$\Omega(J)$,
then the square mass of the system with orbital excitation $J$ and
radial
excitation $n$(0,1,\ldots) is given by (see Ref.~\cite{olss97})
\begin{equation}
\label{m2jn}
M^2(J,n) = M_0^2(J) + M_0(J)\Omega(J)(2n+1).
\end{equation}
The
harmonic approximation is relevant only if it can be assumed that
$\Omega(J) \ll M_0(J)$. We will see in the
next section that the last formula will be
naturally obtained in the light meson case studied in this paper.

\subsection{Application to mesons}
\label{ssec:mesons}

As it is done in previous works, this formalism is applied to the
study
of mesons. The
long range part of the interaction between a quark and an antiquark is
dominated by the confinement which is assumed to be a linear function
of $r$. As its Lorentz structure is not yet determined, we assume that
the confinement is partly scalar and partly vector. The importance of
each part is fixed by a mixing parameter $f$ whose value is 0 for a
pure vector and 1 for a pure scalar. The short range part of the
interaction is assumed to be of vector-type and given by the usual
coulomb-like potential. Thus we have
\begin{equation}
\label{sr}
S(r)=f\,a\,r,
\end{equation}
in which $a$ is the usual string tension, whose value should be
around 0.2 GeV$^2$, and
\begin{equation}
\label{vr}
V(r)=(1-f)\,a\,r-\frac{\kappa}{r}
\end{equation}
in which $\kappa$ is proportional to the strong coupling constant
$\alpha_s$. A reasonable value of $\kappa$ should be in the range 0.1
to 0.6. The quark masses appearing in a spinless Salpeter equation are
the constituent masses. Their values are model dependent, but
generally we have $m_u = m_d \approx 0.2$ GeV and $m_s \approx 0.5$
GeV.

\par We are interested only in Regge trajectories which
give
the behavior of $M^2$ in term of $J$ for large value of $J$. So to
compute the
square mass formula for the particular interaction
(\ref{sr})-(\ref{vr}), we
will use the same technique as in our previous study. A detailed
explanation is given in
Ref.~\cite{silv98}. For a given set of parameters $f$ and $\beta$,
we have to compute the value of $r_0$ as a function of $J$, in
the limit of great values of $J$. In the general case, this radius is
given by a polynomial
equation of 20th degree. This equation is by no mean obvious to obtain
and cannot be considered as a simple extension of the two equal mass
case. The software Mathematica has been used to perform the longest
calculations. The general formulas are very complicated and too long
to be given in detail here, but they can be obtained on request.

\par When $\beta=1$, the old formalism must be recovered. This is not
trivial to demonstrate, but we have verified this point. In
particular, when the two particles have the same mass, the degree of
the general
polynomial is reduced and the solution is one of the root of a second
degree equation. In this case, the determination of $r_0$ and all
subsequent calculations can be performed analytically and are
identical of the ones performed in Ref.~\cite{silv98}.

\par When $\beta$ is different from 1, a numerical solution for $r_0$
must be computed numerically for each value of parameters $f$
and $\beta$. Note that when $f=0$, only one relevant solution ($r_0$
real positive) exists (see Ref.~\cite{silv98}). To find the physical
relevant value of $r_0$, we
generate the solutions for increasing values of $f$ from the known
solution at $f=0$, in order to be sure that the solution is
a continuous function of $f$. This procedure is repeated for each
value of $\beta$. Once the radius is calculated, its value is used to
compute the square meson mass.

\par Finally, we obtain the square mass of the meson under the
following form
\begin{eqnarray}
{M}^2&=&a\,A(f,\beta)\, J+B(f,\beta)\, m\, \sqrt{a\, J}
+C(f,\beta)\,m^2 \nonumber \\
\label{m2}
&+&a\, D(f,\beta) \kappa +a\ E(f,\beta) (2n+1) + O(J^{-1/2}),
\end{eqnarray}
where $0 \leq f \leq 1$, $0 \leq \beta \leq 1$, and $m$ is the mass of
the heaviest quark. The values of
coefficients $A$ and $E$ as a function of
the mixing parameter $f$ and the asymmetry parameter $\beta$ are given
in Figs.~\ref{fig:a}. All coefficients are discussed in the next
section. It is worth noting that the term in $\Omega^2$
naturally does not appear as it is of higher order.

\par When $\beta=1$, that is to say $m_1=m_2=m$, formula (\ref{m2})
reduces to
formula~(20) of Ref.~\cite{silv98}, and coefficients $A$, $B$, $C$,
$D$ and $E$ are given by analytical expressions which can be found in
this reference. We checked theoretically as well as numerically that
our new expressions tend towards the previous ones at the limit
$\beta \rightarrow 1$.

\section{Discussion of the model}
\label{sec:disc}

Equation~(\ref{m2}) is the square mass of a meson for large value of
$J$. Though coefficients of the formula cannot be obtained
analytically, they are calculated for $0 \leq f \leq 1$ and for
$0 \leq \beta \leq 1$, which covers all possible physical situations
provided the mass $m$ appearing in the formula is the mass of the
heaviest quark. One can make the following comments.

\par
\begin{figure}
\includegraphics*[height=8cm]{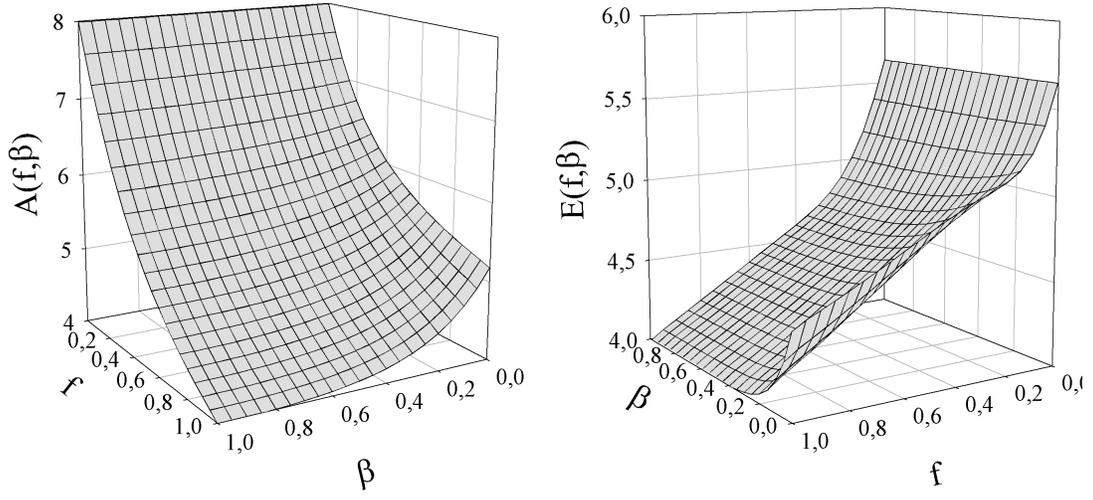}
\vspace{8cm}
\protect\caption{Coefficient $A(f,\beta)$ and $E(f,\beta)$ of
formula~(\ref{m2}) as a
function of the mixing parameter $f$ and the mass asymmetry parameter
$\beta$.}
\label{fig:a}
\end{figure}

 For large value of $J$, the dominant term of the square mass
formula is linear in $J$. This explains the existence of linear Regge
trajectories. The slope of the Regge trajectories depend on three
parameters: the string tension $a$, the mixing parameter $f$ and the
mass asymmetry parameter $\beta$. The string tension is a constant and
gives the energy scale of the slope. The slope depends strongly on the
value of $f$, which is assumed to be also an universal quantity
independent of the system. Unfortunately, as we have seen in our
previous work \cite{silv98}, experimental data from symmetrical mesons
($\rho$ and $\phi$ families) cannot provide reliable information about
the value of $f$. The calculations favor slightly a dominant
vector-type confining interaction ($f=0$). But this work shows that
the slope depends
also on the parameter $\beta$ whose values is fixed by the system. On
Fig.~\ref{fig:a}, one can see that the $\beta$-dependence of the
coefficient $A(f,\beta)$ is only marked when $f \gtrsim 0.5$ and
$\beta \lesssim 0.5$. It is also shown in our previous work
\cite{silv98} that the conclusions we can obtain from our square
mass
formula can be applied to light mesons only, that is to say mesons
containing $u$, $d$ or $s$ quarks. In this case, the only physical
relevant
value for $\beta$ is given by the ratio $m_u/m_s$. For constituent
masses, the value of this ratio must be around 0.5. This shows that
the slope of the Regge trajectory for $K^*$ mesons cannot be
sensitively
different from the ones for $\rho$ and $\phi$ mesons. This is
specially verified if the confinement potential is dominantly of
vector-type ($f \approx 0$) as it seems favored from experimental
data.

\par One can see on Fig.~\ref{fig:a} that the slope of the vibrational
trajectories proportional to $E(f,\beta)$ presents the same
characteristics as
the coefficient $A(f,\beta)$. We can conclude that the slope of the
vibrational trajectory for $K^*$ mesons cannot be sensitively
different
from the ones for $\rho$ and $\phi$ mesons. Obviously, the ratio
$R(f,\beta) = 2\, E(f,\beta)/A(f,\beta)$ is characterized by a similar
behavior as a function of $f$ and $\beta$.

\par A term proportional to $\sqrt{J}$ deforms the Regge trajectories
for low value of $J$. This term vanishes if $m=0$, that is to say the
two quarks are massless, or if $\beta=0$,
that is to say the lightest quark is massless. So, the linearity of
the Regge trajectory is perfect if one quark is massless. This is also
the case if the confinement potential is a pure vector-type
interaction ($f=0$) whatever the masses are. The value
of $B(f,\beta)$ decreases rapidly with $f$ and $\beta$ value
parameters. This fact implies that the linearity of the Regge
trajectory is
only broken in the case of dominantly scalar-type confinement
interaction. As it is mentioned above, this situation is not the one
favored by experimental data. Note that the term proportional to
$\sqrt{J}$ is also independent of the strong coupling constant
$\kappa$.

\par The position of the trajectories with respect to the energy axe
stems from three contributions; one $C(f,\beta)m^2$ is due to the
finite mass the quarks, another negative one $a\,D(f,\beta)\kappa$ is
due to the
strong coupling constant and the last one $a\,E(f,\beta)$ reflects the
zero point motion of the harmonic vibration. These terms are of minor
importance since the zero point energy of the orbital motion cannot
obviously be calculated in our model. Nevertheless it is interesting
to discuss a little bit the behavior of coefficients $C$ and $D$.
The coefficient $C(f,\beta)$
depends clearly on $\beta$ whatever the value of $f$. When $f=0$, this
coefficient can be calculated exactly:
$B(0,\beta)=4 \left( 1 + \beta^2 \right)$.
This term is linked to the contribution of the quark masses to the
meson total mass.
The presence of the Coulomb-like potential decreases the mass of
the meson but has no influence on the Regge trajectories, as it is
expected for a potential which becomes very weak with respect to the
confinement interaction as quark interdistance increases. Again the
coefficient $D(f,\beta)$ is only
affected by the mass asymmetry for $f \gtrsim 0.5$ and
$\beta \lesssim 0.5$.

\par Finally, let us remark that there is no coupling between orbital
and radial motion for large $J$ values (absence of terms $n\,J$). This
is only a consequence of the Coulomb+linear nature of the
quark-antiquark potential. This may not be true for other types of
potentials.

\section{Concluding remarks}
\label{sec:rem}

We have shown that all light mesons exhibit
linear orbital and radial trajectories in the dominantly orbital
states (DOS) model. Slopes of both type of trajectories depend only on
the string tension, the vector-scalar mixture in the confinement
potential and the mass asymmetry between the quark and the antiquark.

\par In the limit of small quark masses, it turns out that the mass
asymmetry between the quark and the antiquark alters sensitively the
slope of the trajectories only for large mass asymmetry and for
dominant
scalar-type confining potential. As the experimental data favor a
vector-type confining potential (see our previous paper \cite{silv98})
and as the mass ratio of non-strange
quark over the strange quark is not very small, the slope of the Regge
and vibrational trajectory for $K^*$ mesons cannot be sensitively
different from the ones for $\rho$ and $\phi$ mesons. This is in
agreement with experimental meson spectra.




\end{document}